\documentclass[useAMS,usenatbib,usegraphicx]{mn2e}

\title[A high--z broad iron line]
 {A broad iron K$\alpha$ line at $z = 1.146$}
\author[A. Comastri, M. Brusa, F. Civano]
  {Andrea Comastri,$^1$
   Marcella Brusa,$^{2,1}$ Francesca Civano$^{2,1}$\\
  $^1$INAF -- Osservatorio Astronomico di Bologna,  
    via Ranzani 1, I--40127 Bologna, Italy\\
    $^2$Dipartimento di Astronomia, Universit\`a di Bologna,
    via Ranzani 1, I--40127 Bologna, Italy\\
}
\date{2004 April 16}

\pagerange{\pageref{firstpage}--\pageref{lastpage}} \pubyear{2004}

\def\LaTeX{L\kern-.36em\raise.3ex\hbox{a}\kern-.15em
    T\kern-.1667em\lower.7ex\hbox{E}\kern-.125emX}

\begin{document}

\label{firstpage}

\maketitle

\begin{abstract}

We report the discovery of a strong iron K$\alpha$ line in the hard X--ray 
selected source {\tt CXOJ 123716.7+621733} in the {\it Chandra} Deep Field 
North survey at $z=$1.146. The analysis is made possible by the very
deep exposure $\sim$ 2 Ms and low background of the {\tt ACIS} detector.
The line profile seems to be inconsistent with a narrow feature. 
The best fit solution is achieved with a broad line.
Most of the flux in the broad component originates at energies 
below 6.4 keV with a  shape similar to that expected from emission 
in the innermost regions of the accretion disk.

\end{abstract}


\section{Introduction}

The strongest emission feature 
in the hard X--ray spectrum of an active galactic nucleus
(AGN) is the fluorescent FeK$\alpha$ emission line at 6.4--7 keV.
Since its discovery by early X--ray observations, it is by now recognized to
be an ubiquitous feature in the high energy spectra of Seyfert 
galaxies. 
The line properties -- in particular the centroid energy, intensity 
and profile -- carry important diagnostic
information about the dynamics and physics of the region where the emission 
originates. 
For example, the best fit line energy is a measure of the ionization status 
of the gas, the line equivalent width (EW) is strongly correlated 
with the amount of fluorescing material, 
while the detection of a broad asymmetric profile, 
first resolved with ASCA by Tanaka et al. (1995) in the bright Seyfert 1 
galaxy MCG--6--30--15, is considered the most direct evidence of the 
presence of an accretion disc extending down to a few gravitational 
radii of the central black hole. 
The power of X--ray spectroscopy as a diagnostic of the AGN physics
is witnessed by hundreds of papers (see Fabian et al. 2000; Reynolds \& 
Nowak 2003, for extensive reviews). 

The detailed study of the 
iron line properties requires a large collecting area and good 
energy resolution. For this reason, most of the results obtained so far 
are limited to nearby, low luminosity Seyfert galaxies 
(e.g. Turner et al. 2002; Fabian et al. 2002; Dewangan, Griffiths \& 
Schurch 2003).  
X--ray observations of sizable samples of higher luminosity AGN, 
extending in the quasars regime ($L_X > 10^{44}$ erg s$^{-1}$), 
reveal the evidence of a trend whereby 
the strength of the iron line decreases and the centroid moves
to higher energies (Iwasawa \& Taniguchi 1993, Nandra et al. 1997a, 
George et al. 2000, Reeves \& Turner 2000, Page et al. 2004).
The search for iron line emission at high redshift is hampered by the 
statistical quality of the X--ray spectra and only a few tentative
detections at the 2$\sigma$ level have been reported.
(e.g., Vignali et al. 1999; Reeves et al. 2001; Norman et al. 2002; 
Gandhi et al. 2004). 

Deep, sensitive X--ray observations carried out with both {\it Chandra}
and XMM--{\it Newton} now offer the opportunity to search for the 
presence of iron emission line with relatively large well defined 
samples of X--ray selected sources, down to limiting fluxes which provides
the possibility to probe a much larger redshift range.
Motivated by this possibility, we have started a systematic search for
emission line features among the serendipitous sources 
in the deepest X--ray observations available to date.

Here we report the detection of a broad iron line
in the X--ray source {\tt CXOJ 123716.7+621733}, serendipitously 
detected in the 2 Ms exposure of the {\it Chandra} Deep Field--North (CDF--N;
Alexander et al. 2003a). 
The spectroscopic redshift ($z=1.146$) is based on a single 
{\tt [OII]$\lambda$3727} line (Barger et al. 2002). 
The compact optical morphology, the hard X--ray spectrum 
and the presence of a radio 
source coincident with the X--ray position
(Richards 2000) leaves no doubt about the AGN nature of this object.
On the basis of the available optical spectrum 
it is not possible to determine if the object is a Type 1 or a 
Type 2 AGN. Unfortunately, 
at the expected {\tt Mg II$\lambda$2798} position the optical 
spectrum is contaminated by strong atmospheric bands and 
night-sky emission lines.
The source is classified as radio--quiet on the basis 
of the 1.4 GHz flux density (0.346 mJy) and optical magnitude 
(Barger et al. 2003).
A cosmological model with $H_0$ = 70 km s$^{-1}$ Mpc$^{-1}$,
$\Omega_m$ = 0.3 and $\Omega_{\Lambda}= 0.7$ is assumed.

\section{X-ray data analysis}

The X--ray data retrieved from the public archive have 
been processed with standard procedures making use 
of the calibrations associated with the {\tt CIAO} 
software\footnote{http://cxc.harvard.edu/ciao/} (version 2.3).
Since the 20 separate {\tt ACIS-I} 
observations that make the 2Ms {\tt CDFN} have been 
performed at different roll angles and aim-points, the source position
in detector coordinates changes during the whole observation. 
Since the {\tt CCD} response depends on the position over the 
detector, source counts have been extracted from each of the 20 observations
taking into account the dependence in size of the PSF with off-axis angle.
The position of our target is always between 3 to 5 arcmin 
from the aim-point, thus the extraction radius was varied from 
4 to 5 arcsec enclosing a constant fraction (90\%) of the PSF.
Background regions were chosen locally for each single observation.
The time dependent quantum efficiency 
degradation\footnote{http://cxc.harvard.edu/cal/Acis/Cal\_prods/qeDeg/}   
of the {\tt ACIS} at low energies was also taken into account 
in the computation of ancillary response files for each dataset.
Spectra, response matrices and effective areas, weighted by the number 
of counts in each observation, were summed using standard 
{\tt FTOOLS} routines.
There is no evidence of substantial flux variability. 
The background subtracted count rates in each of the 20 single observations 
never deviate from the average  ($\sim$ 1.4 counts/ksec) 
by more than 20\% with the exception 
of three exposures where the variation is of the order of 50\%.
The resulting effective exposure time of 1840 ks and the 
net source counts (2620 in the 0.6--7 keV band) are fully consistent 
with the values quoted in the 2Ms X--ray catalog (Alexander et al. 2003a).
The slightly non--standard energy range considered 
for the spectral analysis is driven by the choiche to keep 
the background level always below the 10\% of the total counts.
The summed spectrum was then rebinned with at least 25 counts per bin.
Spectral analysis was carried out with {\tt XSPEC} (Version 11.2), 
errors are reported at the 90\% confidence level for one interesting 
parameter ($\Delta\chi^2$= 2.71).

\subsection{Spectral analysis} 

A single power law fit plus absorption (first line of Table 1)
provides a  
good description of the broad band 0.6--7 keV continuum.
The power law slope is rather flat ($\Gamma \simeq 1.5$) while the 
best fit column density ($\sim$ 1.8$\times$ 10$^{22}$ cm$^{-2}$ 
rest--frame) is clearly inconsistent with
the Galactic column density towards the CDF--N ($N_{Hgal}$ = 1.6 $\times$ 
10$^{20}$ cm$^{-2}$).
The observed 2--10 keV flux of  $\sim$ 1.7 $\times 10^{-14}$ erg cm$^{-2}$ 
s$^{-1}$ corresponds to a rest frame luminosity of 
$\sim$ 7.8 $\times 10^{43}$ erg s$^{-1}$ which is typical of 
a bright Seyfert 1 galaxy. The broad band 0.5--10 keV unabsorbed
luminosity is $\sim$ 1.2 $\times 10^{44}$ erg s$^{-1}$.

Significant residuals around 3 keV are clearly evident  
(Fig.~1), strongly indicating the presence of a line--like feature.

\begin{figure}
 \includegraphics[angle=-90,width=84mm]{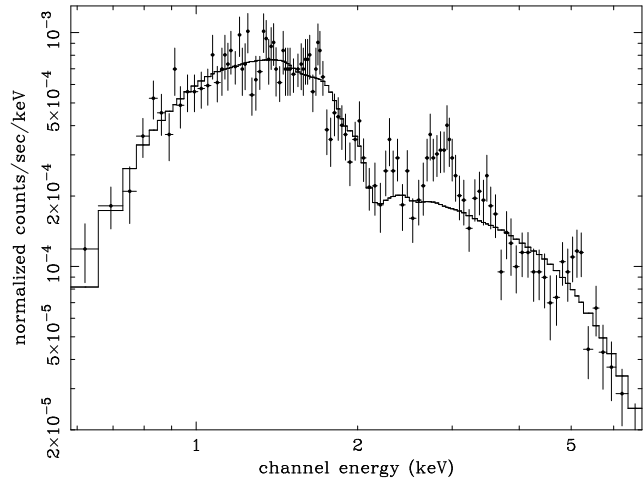}
 \caption{The source spectrum rebinned with at least 25 counts per 
channel is fitted with a single power law ($\Gamma=1.48$) plus 
intrinsic absorption $N_H \simeq 1.8 \times 10^{22}$.}
 \label{sample}
\end{figure}

We therefore added a Gaussian line with rest frame energy fixed at 
6.4 keV. Leaving both the line width and redshift free to vary 
the fit quality is significantly improved (at more than 99.99999\% 
level according to the F--test) and statistically acceptable
(second line in Table ~1). 
The line width is broader than the instrument resolution. 
Moreover, the best fit energy centroid is significantly higher
than that expected on the basis of the spectroscopically measured 
redshift (see Fig.~2). The discrepancy would be even higher 
if a ionised line at 6.7 or 6.96 keV is considered.

\begin{figure}
 \includegraphics[angle=-90,width=84mm]{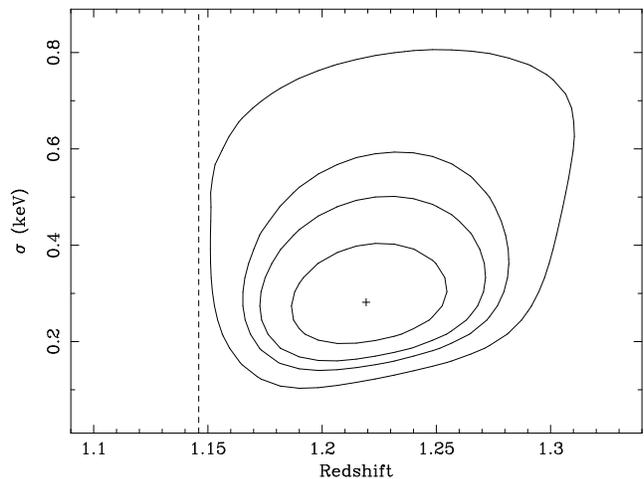}
 \caption{68\%, 90\%, 95\% and 99\% confidence contours for the 
line redshift and Gaussian width $\sigma$. The vertical dashed line
corresponds to the spectroscopic redshift.}
 \label{sample}
\end{figure}

Given that the redshift obtained from optical spectroscopic observations
is unlikely to be affected by such a large error, we have fixed the centroid
of the neutral line at $z=1.146$ and repeated the fitting procedure.
Leaving the line width free to vary, the quality of the fit is 
marginally worse than in the previous case (third line in Table~1). 
From a visual inspection of the spectrum 
it appears evident that most of the line width has to be ascribed 
to a red tail with respect to the 
redshifted centroid of 6.4 keV.  
This effect is clearly seen in Figure~3 where the line width is fixed at the
instrumental energy resolution.

\begin{figure}
 \includegraphics[angle=-90,width=84mm]{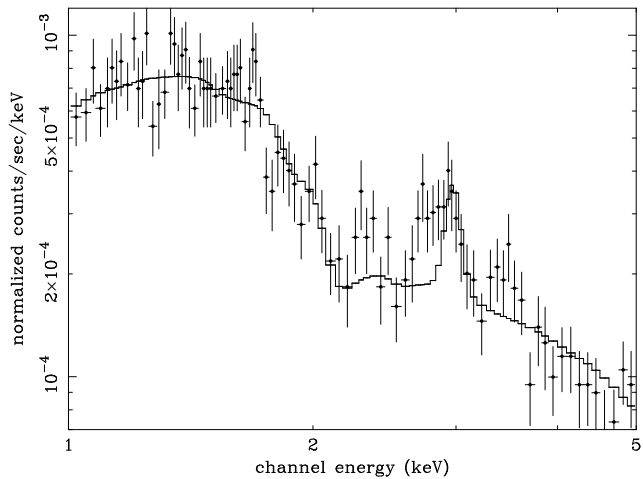}
 \caption{A zoom of the 1--5 keV spectrum fitted with a narrow  
6.4 keV line at the source redshift $z=1.146$. Spectral binning and 
continuum spectrum parameters as in Fig.~1.}
 \label{sample}
\end{figure}

\begin{table*}
\caption{Spectral fit parameters}

\vglue0.3truecm
{\hfill\begin{tabular}{l l l l l l l}
 Model &  $\Gamma$ & $N^a_H$  & $z_{line}$ 
& $\sigma$ (keV) & EW (eV) & $\chi^2$/d.o.f. \\
\hline
Absorbed power law & 1.48$\pm$0.07 & 1.8$\pm$0.3 & ... & ... & ... 
& 134.6/110 \\
&&\\
Absorbed power law + K$\alpha$ line & 1.50$\pm$0.08 & 1.7$\pm$0.3 
& 1.20$^{+0.06}_{-0.02}$ & 0.28$^{+0.15}_{-0.10}$ & 350$^{+109}_{-103}$ 
& 94.8/107 \\
&&\\
Absorbed power law + K$\alpha$ line & 1.52$\pm$0.09 & 1.7$\pm$0.3 
& 1.146(f) & $0.6^{+0.3}_{-0.4}$  & 434$^{+179}_{-193}$ & 104.4/108 \\
\hline
\end{tabular}\hfill}

$^a$ units of $10^{22}$ cm$^{-2}$. \\

(f) fixed parameter \\

\end{table*}

\subsection{Safety checks}

Although the statistical significance regarding the presence
of a line feature at 2.9 keV 
is very robust, as indicated by the F--test value, further checks
have been performed to assess the overall strength of our results. 

First, background spectra have been extracted from several 
regions with different sizes and shapes. In all the cases the background
level is always negligible over most of the considered energy range.
In particular, in the 2.5--3 keV band, where the line feature is 
present, the average background level never exceed 5--6\%  of the
total counts.

\par
Then, to rule out the possibility that something unusual
has occured combining the 20 individual exposures, source spectra have been
extracted for three different time intervals in such a way that each
summed spectrum contains the same number of counts (about 800--1000).
A clear line--like feature is always present around 2.9 keV,
though with a lower statistical 
significance owing to the reduced counting statistic.

\par
Finally, 
to further assess the statistical quality of our results, 
extensive Monte Carlo simulation (up to 100,000 trials) have 
been carried out using the {\tt fakeit} routine within {\tt XSPEC}
(see Alexander et al. 2003b for a similar approach). 
The input model was an absorbed power law spectrum with 
the best fit parameters determined from the observed data (first row 
of Table~1). 
The simulated spectra were then fitted adding a redshifted 
broad Gaussian line with a centroid energy fixed at 6.4 keV rest--frame.
The line intensity, width and redshift are free to vary.
For each simulated spectrum the $\Delta\chi^2$ value
obtained by adding three more free parameters has been computed.
In none of the 100,000 simulations the $\Delta\chi^2$ value is 
as large as that obtained from the fit to the real data.
More specifically, the largest $\Delta\chi^2$ 
of the simulation with 100,000 trials is about 18 to be compared with 
$\Delta\chi^2 \simeq$ 40 as obtained from the observed spectrum (Table~1). 
It is concluded that the probability to detect such a line feature 
by chance is negligible, in good agreement with the 
F--test results.

\subsection{A relativistic line}

The overall properties of the line profile are, at least
qualitatively, similar to those of an emission line originating 
in the innermost region of an accretion disc. 
We next fitted the data with the Schwarzschild disk line model of
Fabian et al. (1989). It is well known that disk models have a considerable
parameter degeneracy and thus the interpretation of the results 
obtained by standard fitting procedures is not straightforward.
Given that the quality of our data is not such to allow a detailed
parameter investigation, the best fit solutions and associated 
error bars have to be treated with caution. 

In all the fits 
we have assumed a neutral iron K$\alpha$ line at 6.4 keV rest frame.
First of all we have considered a model where only the 
inner disc radius is fixed at the last stable orbit of 
a non--rotating black hole (6$R_S$). 
The quality of the fit is statistically acceptable ($\chi^2/d.o.f.$=94/106)
and the improvement with respect to a single power law is significant at
more than 99.9999\% level according to an F--test. 
The formal best fit solution returns a power law index 
for the disk emissivity (which scales as $R^{-q}$) $q \sim 2.5$  
and an inclination angle of 7 degree, while the outer disk radius 
($R_{out}$) is basically unconstrained towards high values. 
The line equivalent width (EW) in the observed frame is $\sim 425\pm125$ eV, 
corresponding to $\sim 910\pm270$ eV in the rest--frame.
The 1--5 keV X--ray spectrum deconvolved by the instrument response
is reported in Figure 4.
Assuming a rather optimistic criterion of  $\Delta\chi^2$=4.6 (corresponding
to 90\% interval for two parameters) we have tried to estimate
errors for both $R_{out}$ and $q$, which, as expected 
are only poorly constrained: $R_{out} > 35 R_S$, $q < 3.5$.

For a disk centrally illuminated by a point X--ray source situated at a
height $H$ above the disk center, $q$ is expected to
lie in the range 0--3 (Fabian et al. 1989) depending on the considered 
disc radius. Most of the line flux originates from radii of the order of $H$ 
where the emissivity index is $q \sim 2$ (Laor 1991). 
Assuming such a value for the disc emissivity law, it is 
possible to better constrain the outer disc radius and inclination
angle (Fig.~5). 
As far as the disc inclination angle is concerned, we note that it 
is always lower than $\sim$ 30 degrees (at a confidence level of
99\%) and rather insensitive to the variation of the other disk line
parameters. Fixing the outer radius to a value of 400$R_S$
the best fit emissivity law and associated 90\% errors 
is $q \simeq 2.3\pm0.6$.

The results above described do not substantially change if the 
continuum spectrum is fitted with a power law plus a reflection component
from cold gas. More specifically a good fit 
($\chi^2/d.o.f.$=97/107; significant at more than 99.9999\% level) 
is obtained when the continuum spectrum
is parameterized by a power law 
with  $\Gamma=$1.8  plus reflection from cold gas subtending a 2$\pi$ sr 
solid angle at the X--ray source. 
The observed equivalent width (EW) with respect to the reflected 
continuum is about 400$\pm$120 eV which 
corresponds to about  860$\pm$260 eV in the source rest frame.

\begin{figure}
 \includegraphics[angle=-90,width=84mm]{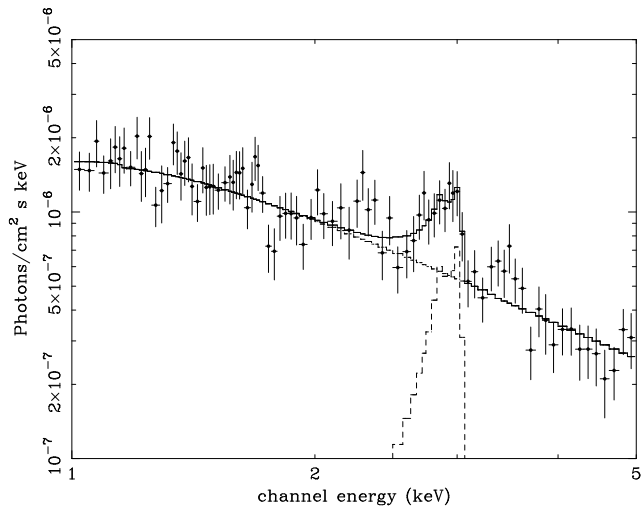}
 \caption{The unfolded spectrum fitted with an absorbed 
power law ($\Gamma \simeq$ 1.5; $N_H \simeq 1.7 \times 10^{22}$ cm$^{-2}$)
plus a relativistic disk line. Spectral binning as in Fig.~1.}
 \label{sample}
\end{figure}

\begin{figure}
 \includegraphics[angle=-90,width=84mm]{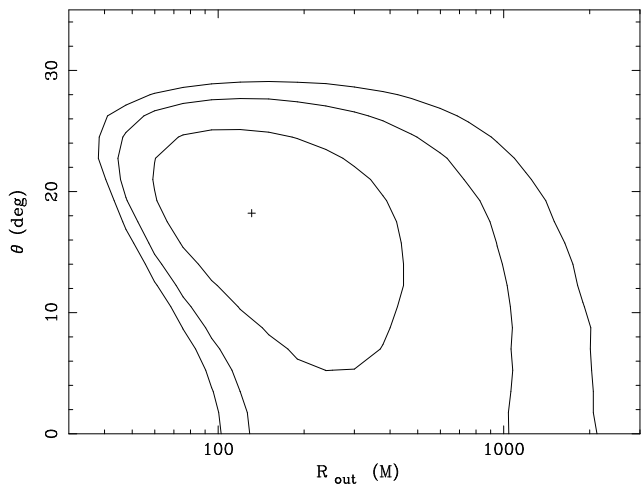}
 \caption{68\%, 90\% and 95\% confidence contours for the 
 disc inclination angle and outer radius $R_{out}$}
 \label{sample}
\end{figure}

\section{Discussion and Conclusions}

A strong emission feature due to the K$\alpha$ iron line
has been clearly detected in the X--ray spectrum of the relatively bright, 
hard X--ray source {\tt CXOJ 123716.7+621733} at $z$=1.146 in the 
{\it Chandra} Deep Field North.
The broad band 0.6--7 keV {\it Chandra} spectrum 
is best described by a relatively hard ($\Gamma \simeq$ 1.5) 
power law plus significant intrinsic absorption 
$N_H \sim 1.8 \times$ 10$^{22}$ cm$^{-2}$.  
An almost equally good 
fit is obtained if the continuum emission is parameterized 
with a steeper ($\Gamma$ = 1.8) slope plus a reflection component 
as commonly observed in nearby Seyfert galaxies.

The line profile is not consistent with a narrow feature,
being significantly broader than the {\tt ACIS-I CCD} resolution
at that energy. Such a result does not depend on the adopted
shape of the underlying continuum.

In order to reproduce both the observed width and the optical 
spectroscopic redshift, most of the line flux has to originate 
in a red wing. An adequate description of the overall line shape 
has been obtained with a relativistic line model.  
Although the quality of the data is not such to break the degeneracy 
between the various line parameters, the present observation suggests
an almost face--on orientation of the accretion disk.

The line equivalent width (860$\pm$260 eV) is larger than the average values 
measured by {\tt ASCA} for a sample of nearby bright Seyfert galaxies.
Nandra et al (1997b) report an average value of 230$\pm$60 eV,
for a relativistic line model.
Plausible possibilities include 
iron overabundance as in MCG--6--30--15 (see e.g. Lee et al. 1999)
and/or an enhanced contribution of the reflected flux possibly associated
to time variability. The latter possibility appears unlikely given 
the lack of significant flux variability over more than two years.
It must be noted that large values of the EW are not unusual among 
relatively faint X--ray sources (George et al. 2000) and likely
to be due to a selection effect associated to the 
uncertainties in modelling the continuum and line spectrum.

The broad band properties of {\tt CXOJ 123716.7+621733} 
(namely, the high X--ray to optical ratio, (log$f_X/f_{opt} \simeq$ 1), 
the X--ray column density ($>$ 10$^{22}$ cm$^{-2}$) and the 
optical, near--infrared colors ($B-V=0.7$, $R-K=4.4$; Barger et al. 2003)) 
indicate substantial obscuration of the nuclear source.
There is also no evidence of typical AGN emission lines 
in the low resolution optical spectrum (Barger et al. 2002).

The present results indicate that the iron K$\alpha$ 
line could be the strongest AGN signature
in the broad band spectra (from optical to X--rays) of distant, 
obscured AGN for which only a low signal--to--noise optical spectrum 
covering a relatively small range of wavelengths is available,
and highlight the uniqueness of X--ray observations
in the study of supermassive black holes at cosmologically
interesting distances. 

The study of relativistically broadened iron line profiles
would greatly benefit from deep X--ray observations 
with the XMM--{\it Newton} large collecting area telescopes.
Observations with the new generation of X--ray observatories 
({\tt Constellation X} and {\tt XEUS}) will allow systematic studies
of iron line properties at high redshifts.

\section{Acknowledgments}

It is a pleasure to thank Piero Ranalli and Cristian Vignali for help
in data reduction and informative discussions and Gianni Zamorani
for a careful reading and suggestions. We also thank the referee,
Dave Alexander, for detailed comments which improved the presentation
of the results. The authors acknowledge partial support by the ASI contract
I/R/057/02, the INAF grant 2003/270 and the MIUR grant Cofin--03--02--23.

\end{document}